\documentclass[a4paper]{article}
\pdfoutput=1
\usepackage{multirow}
\usepackage{algorithm, algorithmic}
\usepackage{ISCSLP2022}

\title{TSUP Speaker Diarization System for Conversational Short-phrase Speaker Diarization Challenge}

\name{Bowen Pang$^1$, Huan Zhao$^1$, Gaosheng Zhang$^2$, Xiaoyue Yang$^2$, Yang Sun$^2$, Li Zhang$^1$, Qing Wang$^1$, Lei Xie$^1$}
\address{
$^1$Audio, Speech and Language Processing Group (ASLP@NPU), School of Computer Science, \\
Northwestern Polytechnical University (NPU), Xi'an, China  \\
  $^2$Shenzhen Transsion Holding Limited}
\email{ 
\{zhaohuan, pangbowen\}@mail.nwpu.edu.cn, \{gaosheng.zhang, xiaoyue.yang, yang.sun\}@transsion.com, lxie@nwpu.edu.cn}

\begin{document}

\maketitle
\begin{abstract}
   This paper describes the TSUP team's  submission to the ISCSLP 2022 conversational short-phrase speaker diarization (CSSD) challenge which particularly focuses on short-phrase conversations with a new evaluation metric called conversational diarization error rate (CDER). In this challenge, we explore three kinds of typical speaker diarization systems, which are spectral clustering~(SC) based diarization, target-speaker voice activity detection~(TS-VAD) and end-to-end neural diarization~(EEND) respectively. Our major findings are summarized as follows. First, the SC approach is more favored over the other two approaches under the new CDER metric. Second, tuning on hyperparameters is essential to CDER for all three types of speaker diarization systems. Specifically, CDER becomes smaller when the length of sub-segments setting longer. Finally, multi-system fusion through DOVER-LAP will worsen the CDER metric on the challenge data. Our submitted SC system eventually ranks the third place in the challenge.
\end{abstract}

\section{Introduction}

Speaker diarization is to determine ``who spoke when" in an audio stream that may contain an unknown number of speakers~\cite{anguera2012speaker,park2022review}. It is an indispensable task in multimedia information retrieval, speaker turn analysis and audio processing~\cite{wang2018speaker,basu2016overview}. In particular, speaker diarization has the potential to significantly improve 
automatic speech recognition~(ASR) accuracy in multi-speaker conversation scenarios~\cite{xiao2021microsoft,yu2022m2met}. 


Clustering-based methods have dominated speaker diarization for many years, which is composed of multiple, independently-optimized modules including voice activity detection~(VAD), speech segmentation, speaker embedding extraction and speaker clustering~\cite{ning2006spectral,lin2019lstm}. Although these systems have shown superior performance in several speaker diarization challenges~\cite{xiao2021microsoft,wang2021scenario}, there are apparently two defects in such clustering-based methods. Specifically, they cannot properly deal with speaker overlap and the independent optimization of different sub-modules may lead to sub-optimal performance~\cite{raj2021multi,wang2021bytedance,park2019auto}.

To deal with the overlapping speech problem particularly, speech separation can be used as a pre-processing step~\cite{xiao2021microsoft} and target-speaker voice activity detection~(TS-VAD) can be adopted as a post-processing step~\cite{medennikov2020target}. 
Particularly, the use of TS-VAD has led the speaker diarization system to achieve state-of-the-art (SOTA) performance in several open challenges~\cite{wang2021scenario,wang2021dku}. However, TS-VAD has a drawback that the oracle information about the maximum number of speakers in an audio stream has to be known in advance.

Recently, end-to-end neural diarization~(EEND)~\cite{fujita2019end, fujita2019endsa} has been proposed to deal with both overlapping speech as well as to directly optimize a diarization system via diarization error minimization. Specifically, EEND treats speaker diarization as a classification task and estimates speech activities of all speakers jointly frame-by-frame. To solve the permutation problem~\cite{adavanne2017report}, Fujita et al. introduced a permutation-free scheme~\cite{hershey2016deep,yu2017permutation} particularly into the training objective function. Thus the EEND system is trained in an end-to-end fashion under the objective function that provides minimal diarization errors.


To advance the speaker diarization performance in conversational short-phrase scenario, ISCSLP2022 specifically held a CSSD challenge with a new conversation dataset. In conversions like agent-customer telephone calls, the speech utterance from each side is usually very brief and sometimes it is limited to only a short phrase, which results in frequent speaker changes. Speaker diarization in such a scenario poses particular challenges to both current systems as well as evaluation metrics. Different from the previous speaker diarization challenges~\cite{ryant2020third,brown2022voxsrc}, the evaluation metric of the CSSD challenge is the so-called \textit{conversational diarization error rate}~(CDER)~\cite{cheng2022conversational} rather than the typical diarization error rate~(DER). Although DER has been used as a standard metric for speaker diarization for a long time, it fails to give enough emphasis to the short conversational phrases which last for a short time but accurate discrimination on them play a vital role to the downstream tasks such as speech recognition and understanding. Different from DER calculated on the time duration level, CDER is regardless of the length of the utterance, and all types of mistakes are equally reflected in the evaluation metric~\cite{cheng2022conversational}. Under this measurement, short-phrase and long sentences have identical importance.


Considering the specific task and the new evaluation metric, in this paper, we make a comparative study on the three typical speaker diairzation approaches -- spectral clustering (SC), TS-VAD and EEND. Our major findings are twofold. First, SC is more preferred by the new CDER metric because frame-level prediction nature of TS-VAD and EEND may lead to more errors on short segments, which makes small impact on DER but significant impact on CDER which equally treats long and short utterances. Second, tuning on hyperparameters is essential to CDER for all three types of speaker diarization system. Specifically, CDER will be reduced when the length of sub-segments setting longer. Finally our submitted spectral clustering system manages to win the third place in the challenge. 

\begin{figure*}[th]
    \centering
    \includegraphics[width=1.0\textwidth]{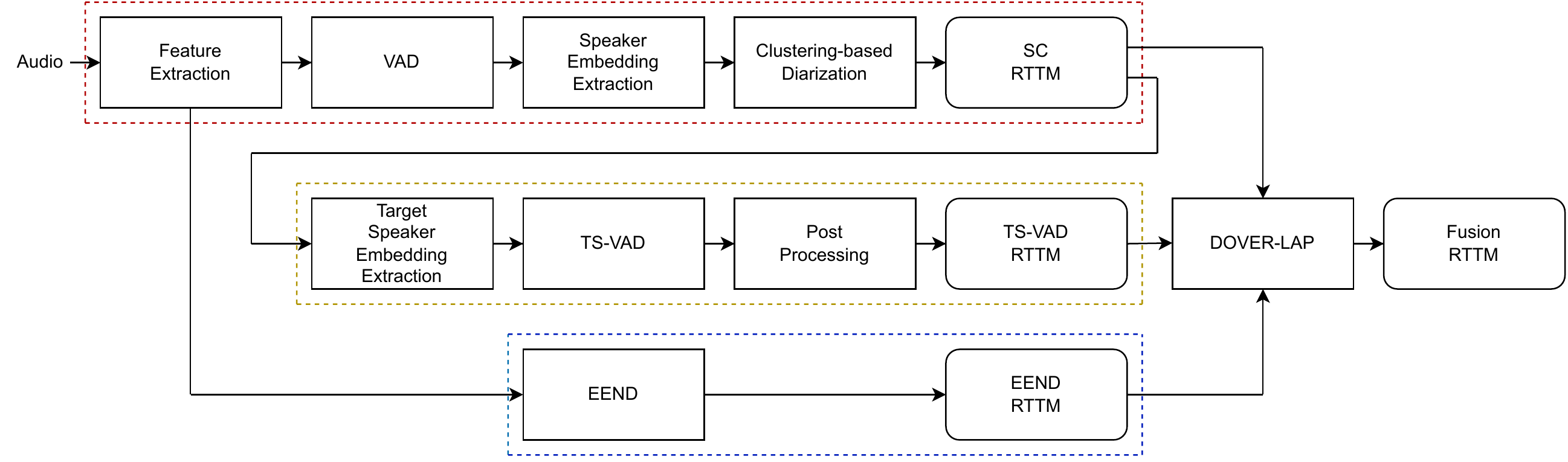}
    \caption{The overview of three kinds of speaker diarization systems and their fusion}
    \label{fig:my_label}
\end{figure*}

\section{System Description}

In the CSSD challenge, we explore several systems including SC, TS-VAD and EEND, which are summarized in Figure~\ref{fig:my_label}. In addition, we also explore DOVER-LAP~\cite{raj2021dover} to fuse the RTTM outputs inferred from the above three systems to get the final speaker diarization results.


\subsection{Spectral Clustering System}

A clustering-based speaker diarization system generally consists of VAD, speaker embedding extractor and clustering modules. In general, the VAD module can be simple energy-based or neural network based. Thanks to the recent advances in neural speaker verificatoin, various types of speaker embedding extractors, including x-vectors~\cite{snyder2018x}, ResNet~\cite{he2016deep} and ECAPA-TDNN~\cite{desplanques2020ecapa}, can be considered to extract discriminative speaker embeddings from audio segments. As for the clustering step, agglomerative hierarchical clustering~(AHC)~\cite{murtagh2012algorithms} and spectral clustering~\cite{von2007tutorial} are the two typical methods.

In this challenge, we implement a ResNet-LSTM based VAD module similar to ~\cite{wang2021dku} using the allowed training data and its performance on the dev and test sets is comparable with that achieved by the TDNN-based VAD in Kaldi~\cite{povey2011kaldi} which is trained using more data. The timestamp labels during training are generated from the transcripts of the training set. In our VAD module, the ResNet structure is first to extract frame-level feature map. Then the current frame feature map is concatenated with the feature maps from both the previous and next frames and the concatenated feature map then goes through a statistics pooling layer. Finally, two BLSTM layers and a linear layer are used to obtain the probabilities of speech for the current frame. 

The structure of our speaker embedding extractor is ResNet34~\cite{wang2022cross}, which is used to extract utterance-level embeddings after speaker segmentation. Finally, we adopt the spectral clustering algorithm to cluster the speaker identity of the embeddings. Spectral clustering has been widely adopted in speaker diarization. The conventional AHC approach is highly time-consuming where the processing time for an audio stream depends on the number of the initial segments. SC is thus introduced to mitigate this problem.



In details, spectral clustering is a graph-based clustering algorithm. After scoring pairs of sub-segment embeddings after speaker segmentation, we can get a similarity matrix. Given the similarity matrix $S$, SC finds a partition of the graph such that different groups edges have very low weights. Our implementation on SC is summarized in Algorithm 1.


\begin{algorithm}
\renewcommand{\algorithmicrequire}{\textbf{Input:}}
\renewcommand{\algorithmicensure}{\textbf{Output:}}
\caption{Algorithm of spectral clustering}
\label{alg:sc}
\begin{algorithmic}[1]
		\REQUIRE The similarity matrix $S \in \mathbb{R}^{n \times n}$
		\ENSURE The clustering label of every segment embedding.
		\STATE Set similarity matrix $S$ diagonal elements to 0.
		\STATE Compute the normalized Laplacian $L_{norm}$ of matrix $S$.
		\STATE Compute the eigenvalues and eigenvectors of $L_{norm}$.
		\STATE Compute the number of cluster $k$. In our implementation, we set a threshold $\alpha$ and count the number of eigenvalues which is below $\alpha$ as $k$.
		\STATE Concatenate the eigenvectors $\lambda_1, \cdots, \lambda_k$ as columns corresponding to the $k$ smallest eigenvalues into a matrix $P \in \mathbb{R}^{n \times k}$.
		\STATE For $i=1,\cdots,n$, let $r_i$ be the $i$-th row of $P$. Cluster row vectors $r_1,\cdots,r_k$ with the $k$-means algorithm.
		\STATE \textbf{return} The label of each embedding $A_1, A_2, \cdots, A_n$ where $A_i=\{j\vert j=1,\cdots,k\}$
\end{algorithmic}  

\end{algorithm}

\subsection{Target-speaker Voice Activity Detection}

A TS-VAD system~\cite{medennikov2020target} aims to handle the overlapped speech in speaker diarization, which adopts the target-speaker embeddings to identify the specific speakers within the overlapped speech. The target-speaker embeddings are estimated based on the initial diarization results from the clustering-based system.

Our TS-VAD system follows the structure in~\cite{wang2022cross}, which consists of a ResNet front-end and a detection back-end. We use a deep speaker embedding model to extract frame-level embeddings instead of the original TS-VAD~\cite{medennikov2020target} taking four CNN layers to process the acoustic features. For an audio, we expect to detect speech segments for each speaker. First, we use a ResNet34 network which has the same structure with the speaker embedding model to extract frame-level embeddings. Unlike speaker embedding model that applies statistics pooling to project variable-length utterances into fixed-length speaker representation embeddings, our TS-VAD employs the same pooling on each frame which is combined with its adjacent frames to obtain frame-level embeddings. Then, the frame-level embeddings are concatenated with the target-speaker embeddings. Next, a Transformer encoder is used to extract detection state of each speaker. After that a BLSTM layer processes these detection states which are concatenated together to find relationship between each speaker. Finally, a linear layer with a sigmoid function are used to determine the probabilities of each speaker at every time step. By analyzing and processing the speech probabilities, we can obtain speech segments of each speaker.



\subsection{End-to-End Neural Diarization}

EEND adopts a single neural network to obtian the final diarization result directly from input audio feature. EEND is composed of an encoder block and a binary classification block. Given a sequence generated from an audio signal, encoder block extracts the feature containing diarization information and classification block estimates a two-dimensional sequence to express a probability of whether a speaker speaks at a frame. We employ permutation invariant training~(PIT)~\cite{fujita2019end} to traverse all orders and optimize the minimal one because changing speakers order does not affect the final result in diarization task. Eventually, each frame is classified into one of the following three cases, non-speaker, one-speaker and overlap.

In this challenge, we attempt two kinds of model structures, which are residual auxiliary EEND~(RX-EEND)~\cite{yu2022auxiliary} and speaker-wise chain EEND~(SC-EEND)~\cite{fujita2020neural}. In RX-EEND, each encoder block is enriched by a residual connection to restrict gradient to a reasonable range. In addition, the output tensor of the encoder block with the exception of the last block is aggregated to calculate an extra auxiliary loss to gather more diarization information. In SC-EEND, there is an LSTM layer to model the relation that whether a speaker spoken in the past will affect other speakers between different speakers since we will indeed be temporarily silent when the other person is talking during the real conversation. RX-EEND improves the structure of the network and SC-EEND models a more appropriate structure in the application scene.

\section{Experiments}
In this section, we first describe the model configuration and training details. Then the score metric used in the challenge is introduced. Finally we report the experimental results and major findings.

\subsection{Model Configuration and Training}

\subsubsection{Speaker Embedding Model}

Following the challenge rules, our experiments are conducted on VoxCeleb~\cite{nagrani2017voxceleb}, CN-Celeb~\cite{fan2020cn} and MagicData-RAMC~\cite{yang2022open} datasets. We use ResNet34 as the speaker embedding model which has \{32, 64, 128, 256\} channels of residual blocks. The pooling layer is statistics pooling and the embedding size is 128. AAMSoftmax~\cite{deng2019arcface} with a margin of 0.2 and scale of 32 is used to train the ResNet34 model. The input of model is 1.5s fix-chunked wave and the acousic feature is 80-dim log Mel filter bank feature with 25ms frame length and 10ms frame shift. Specifically, we use a two-stage training strategy to optimize the speaker embedding extractor. First, we pre-train the speaker embedding extractor for 10 epochs with the VoxCeleb and CN-Celeb datasets and the learning rate is set to 0.01. To mitigate the mismatch between the pre-training datasets and the MagicData-RAMC data, we adopt fine-tuning to get the in-domain speaker embedding extractor. Specifcially, the model is fine-tuned for 30 epochs using the non-overlapped speech segments of MagicData-RAMC training set and the learning rate is set to 0.001. The optimizer is Adam. We also perform online data augmentation during the training of the speaker embedding model. Specifically, frequency-domain SpecAug~\cite{park2019specaugment}, additive noise augmentation~\cite{snyder2015musan} and reverberation augmentation~\cite{habets2006room} are adopted.





\subsubsection{VAD}
We implement a ResNet-LSTM based VAD model which follows the configuration in~\cite{wang2021dku}. The input is 16s chunked waveform, and the acoustic feature is 80-dim log Mel filter bank with a frame size of 25ms and a frame shift of 10ms. The model is trained on the MagicData-RAMC training set for 30 epochs with a learning rate of 0.0001. We optimize the model with binary cross-entropy (BCE) loss and Adam optimizer.

\subsubsection{TS-VAD}

We build a simulated dataset from the MagicData-RAMC training set for training the TS-VAD model. The data simulation method is similar to that described in~\cite{wang2022cross}. First, we extract all non-overlapped speech segments from the MagicData-RAMC training set for each speaker. Then, the corresponding labels are extracted from the transcripts associated with the MagicData-RAMC training set. The difference is that we do not remove the silence region, which aims to learn silence information so that TS-VAD can detect speaker change region better. Finally, in training stage, we fill the activated region with non-overlapped speech segments.


The ResNet front-end configuration is the same as the speaker embedding model as well as the Transformer encoder includes four encoder layers with 256 attention units containing 2 heads and 1024 internal units. The TS-VAD model is first trained on the simulated dataset for 10 epochs with freezed ResNet34 front-end. Next, we train the whole system for 5 epochs further on the simulated dataset. Finally, we fine-tune the whole system using the MagicData-RAMC training set. We apply the learning rate of 0.0001 when training on the simulated dataset and 0.00001 during the fine-tuning stage respectively. The model is optimized with binary cross-entroy~(BCE) and Adam optimizer.

In the inference stage of TS-VAD model, we first extract all the non-overlapped speech segments from the results of the SC system. Then we extract the target-speaker embeddings. Next, we split one audio stream into 16s chunk waves with a 4s shift as the input of the TS-VAD model. After getting the probabilities of each speaker at the frame level, we use median filtering with a window size of 5 to smooth the probabilities. Then we binarize the probabilities with the threshold of 0.9 and delete speech segments shorter than 0.1s.

\subsubsection{EEND}

In the RX-EEND and SC-EEND models, we use four encoder blocks with 256 attention units containing four heads. We use 2048 internal units in a position-wise feed-forward layer. The input feature is 23-dim log Mel filter bank with total of 15 frames context and sub-sampling ratio is 10. The loss function is BCE with the permutation-free scheme. The Adam optimizer is used, where the learning rate schedule with warm-up steps of 100,000 is applied~\cite{vaswani2017attention}.

EEND has two stages in training. Initially, we simulate 100,000 two-speakers conversations as training data and reserve 500 for evaluation. The simulation algorithm is described in~\cite{fujita2019end}. We train a pre-train model using these simu-data for 100 epochs. Afterwards, we fine-tune the pre-train model using MagicData-RAMC training set for 30 epochs with 0.00001 learning-rate. In the inference stage, we prepare a 50 frames buffer~\cite{xue2021online} to save partial input sequence and corresponding prediction from model to prevent memory overflow, because the average duration of MagicData is 30 minutes which is too large to process the entire audio in GPU.

\subsection{Score Metric}

The speaker embedding models are evaluated by equal error rate~(EER) and min detection cost function~(minDCF) with $P_{target} = 0.01$ and $C_{FA}=C_{Miss}=1$.

The speaker diarization task generally employs diarization error rate~(DER) as an evaluation metric on time duration level. In order to make all types of mistakes equally reflect in the final evaluation metric, this challenge adopts conversational-DER~(CDER)~\cite{cheng2022conversational} to evaluate the performance of the speaker diarization system on the sentence level.


\subsection{Experimental Results}

\subsubsection{Speaker Embedding Model}
Speaker embedding extractor contributes a lot to the performance of the speaker diarization system. So we first evaluate the EER/minDCF of our speaker embedding extractor and results which are shown in Table~\ref{tab:sv-results}. After fine-tuning, the EER/minDCF on MagicData-RAMC dev and test set are absolutely reduced by 0.45\%/0.04 and 0.17\%/0.02 respectively.
\begin{table}[th]
\centering
\captionsetup{font={footnotesize}} 
\caption{The EER~(\%) and $minDCF_{0.01}$ of speaker embedding model.}
\label{tab:sv-results}
\resizebox{\linewidth}{!}{
\begin{tabular}{ccccccccc}
\toprule
\centering
\multirow{2}{*}{\text{Training}}  & \multicolumn{2}{c}{Vox-O} & \multicolumn{2}{c}{CN-Eval} & \multicolumn{2}{c}{MagicData-Dev} &
\multicolumn{2}{c}{MagicData-Test}
\\ \cmidrule{2-9} 
                    & EER           & minDCF      & EER            &minDCF       & EER             & minDCF       & EER            & minDCF          \\ \midrule
Pre-training                     & 1.27         & 0.14        & 7.13          & 0.45          & 3.40          & 0.56          & 3.32          & 0.26         \\
 Fine-tuning                     & -              & -             & -               & -               & \textbf{2.95}          & \textbf{0.52}          & \textbf{3.15}           & \textbf{0.24}          \\ \bottomrule
\end{tabular}
}\vspace{-8pt}
\end{table}

\subsubsection{VAD}
Table~\ref{tab:sad-results1} shows the false alarm (FA), miss detection (MISS) and accuracy(ACC) on MagicData-RAMC dev set and test set. We report the performance of Kaldi VAD model and our ResNet-LSTM VAD model at different thresholds. As presented in Table~\ref{tab:sad-results1}, the ResNet-LSTM VAD model and Kaldi VAD model have comparable performance. When we evaluate CDER performance after each VAD model is integrated into the overall speaker diarization system, we find ResNet-LSTM VAD(0.5) model achieves the lowest CDER. Therefore, ResNet-LSTM is selected as the VAD module in our submitted speaker diarization system.

\begin{table}[th]
\captionsetup{font={footnotesize}} 
\caption{The false alarm (FA), miss detection (MISS) and accuracy~(\%) of VAD models at different threshold.}
\label{tab:sad-results1}
\resizebox{\linewidth}{!}{
\begin{tabular}{ccccccc}
\toprule
\multirow{2}{*}{\text{Model}} & \multicolumn{3}{c}{\text{MagicData-Dev}}                & \multicolumn{3}{c}{\text{MagicData-Test}}               \\ \cmidrule{2-7} 
                                & \text{FA} & \text{MISS} & \text{ACC} & \text{FA} & \text{MISS} & \text{ACC} \\ \midrule
Kaldi VAD~\cite{povey2011kaldi}   & 8.19             & \textbf{4.31}               & \textbf{95.69}                  & 8.35             & \textbf{4.84}               & \textbf{95.16}                  \\
ResNet-LSTM~(0.5)      & \textbf{0.37}             & 7.89               & 92.11                  & \textbf{0.80}             & 6.95               & 93.05                  \\
ResNet-LSTM~(0.4)
      & 0.46             & 6.87               & 93.13                  & 0.96             & 5.95               & 94.05                  \\
ResNet-LSTM~(0.3)      & 0.58             & 5.98               & 94.02                  & 1.14             & 5.00               & 95.00                  \\ \bottomrule
\end{tabular}
}
\vspace{-10pt}
\end{table}

\subsubsection{Speaker Diarization}
Table~\ref{tab:dia-results1} shows the DER~(collar=0.25) and CDER results of all the systems on the MagicData-RAMC dev, test set and final blind test set. Results show that all the performance of our models are better than those of baseline~\cite{cheng2022conversational} on CDER. For the TS-VAD model, we obtain the best CDER of 9.70\% on the dev set. In addition, the SC system achieves the lowest CDER of 9.50\% on the test set. On blind-test set, the TS-VAD model does not achieve the acceptable result. Finally, we obtain a CDER of 9.10\% in SC system, which is our submitted results on the leadboard of CSSD challenge.
\begin{table}[th]
\centering
\captionsetup{font={footnotesize}}
\caption{The DER~(\%) and CDER~(\%) of different speaker diarization systems on MagicData-RAMC dev, test and binld test sets.}
\label{tab:dia-results1}
\resizebox{\linewidth}{!}{
\begin{tabular}{cccccc}
\toprule
\multirow{2}{*}{\text{Model}} & \multicolumn{2}{c}{\text{MagicData-Dev}}            & \multicolumn{2}{c}{\text{MagicData-Test}}           & \text{Blind-Test} \\ \cmidrule{2-6} 
                       & \text{DER}   & \text{CDER}          & \text{DER}  & \text{CDER}          & \text{CDER}          \\ \midrule
Baseline               & \textbf{5.57}             & 26.90          & \textbf{7.96}           & 28.20          & -          \\
SC                     & 14.33           & 12.00          & 18.19           & \textbf{9.50} & \textbf{9.10}       \\
TS-VAD                 & 12.79          & \textbf{9.70}  & 17.52           & 10.50         & 16.40          \\
RX-EEND                & 12.05           & 16.30          & 12.01           & 19.20          & -          \\
SC-EEND                & 11.29           & 17.80          & 13.36           & 24.20          & -          \\ \midrule
SC\&TS-VAD Fusion & 18.22 & 11.40 & 16.82 & 13.40 & - \\
 \bottomrule
\end{tabular}
}\vspace{-8pt}
\end{table}

\begin{figure}[th]
    \centering
    \includegraphics[width=0.9\linewidth]{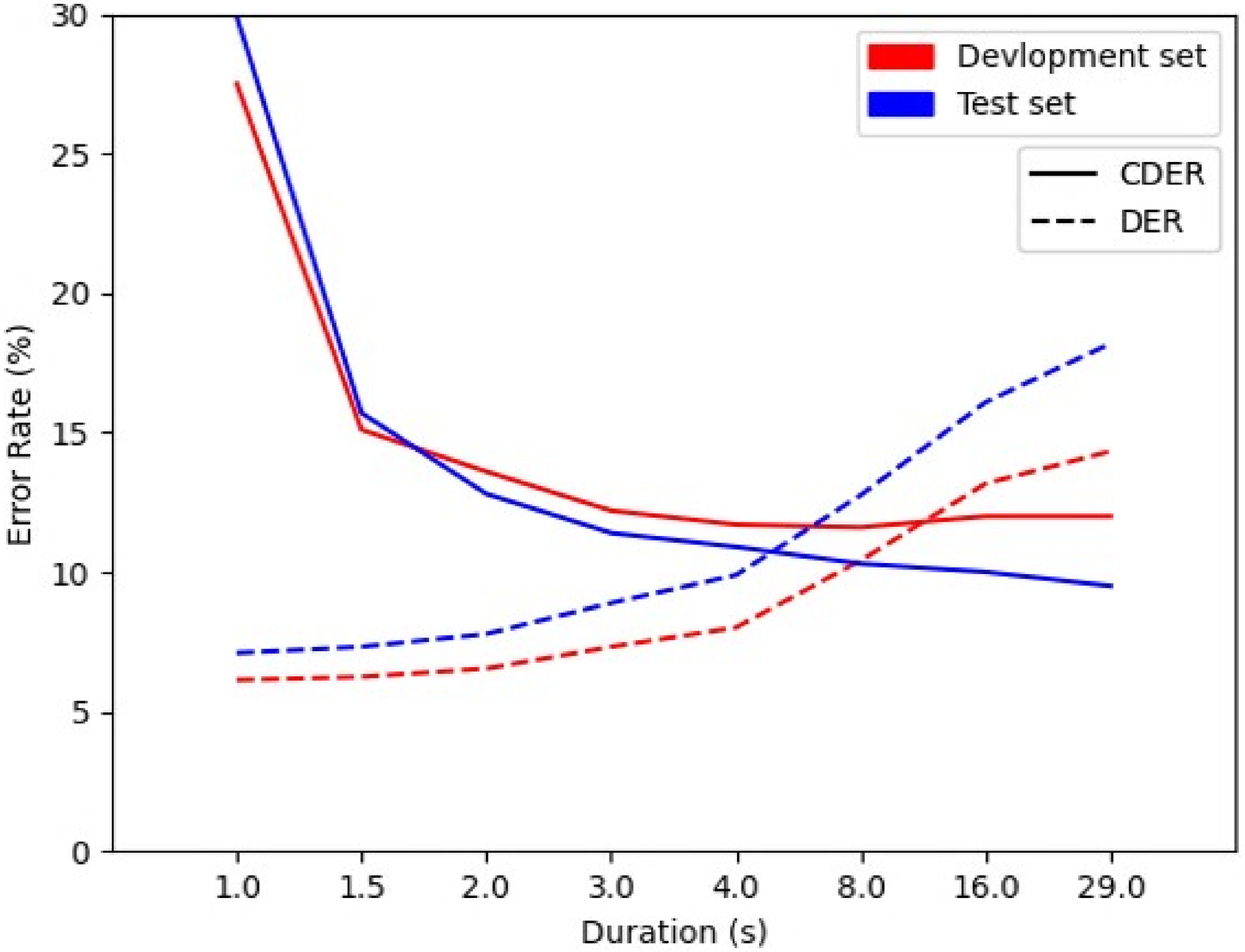}\vspace{-10pt}
    \caption{The influence of different sub-segment duration to DER~(\%) and CDER~(\%)}
    \label{fig:durations}\vspace{-5pt}
\end{figure}

After manual tuning hyperparameters, we find that CDER is reduced as the length of sub-segments setting longer. Figure~\ref{fig:durations} shows the trend of the influence of sub-segments duration to the DER/CDER. Since the penalty of mistakes caused by different lengths in the calculation of CDER is the same, the expanse of the mistake penalty caused by short sub-segments is relatively larger. We observe that longer sub-segment duration can help to reduce this kind of mistake, which improves the performance of CDER at the same time. This means that a sentence from one speaker is divided into several sub-segments with short duration, and these sub-segments may be recognized as two or more speakers, which leads to serious error. As a result, longer sub-segment can help to avoid this problem.

Furthermore, the DOVER-LAP~\cite{raj2021dover} fusion strategy is used to fuse the SC and TS-VAD systems. However, the CDER of the fusion result decreases to 11.40\%/13.40\% on MagicData-RAMC dev and test sets respectively. This is mainly because there is few speaker overlaps in the MagicData-RAMC dataset while DOVER-LAP is originally designed to better handle overlapped speech via voting.

\section{Conclusion}

In the ISCSLP2022 CSSD challenge, we have explored three different speaker diarization systems -- spectral clustering (SC), target speaker VAD (TS-VAD) as well as end-to-end neural diarization (EEND). Our study shows that spectral clustering based speaker diarization is still competitive in the current challenge setup, i.e., accessing speaker diarization performance via the new CDER metric for conversations composed of short-phrases. With the submitted SC system, we achieve our lowest CDER of 12.0$\%$ and 9.5$\%$ on the dev set and test set respectively, eventually leading our system to the 3rd rank on the blind test set.

\end{document}